\documentclass[12pt]{iopart}
\usepackage{graphicx}
\usepackage{wrapfig}
\usepackage{siunitx}
\usepackage{amsfonts}
\usepackage{url}
\usepackage{xcolor}
\expandafter\let\csname equation*\endcsname\relax
\expandafter\let\csname endequation*\endcsname\relax
\usepackage{amsmath}
\begin{document}
\title[Krebs et al. -- VDE benchmark of nonlinear MHD codes]{Axisymmetric simulations of vertical displacement events in tokamaks: A benchmark of M3D-C$^1$, NIMROD and JOREK}%A benchmark of nonlinear MHD codes
\author{I~Krebs$^{1,2}$, F~J~Artola$^3$, C~R~Sovinec$^4$, S~C~Jardin$^1$, K~J~Bunkers$^4$, M~Hoelzl$^5$, N~M~Ferraro$^1$}
\address{$^1$ Princeton Plasma Physics Laboratory, P.O. Box 451, Princeton, New Jersey 08543-0451, USA}
\address{$^2$ FOM Institute DIFFER - Dutch Institute for Fundamental Energy Research, P.O. Box 6336, 5600 HH Eindhoven, The Netherlands}
\address{$^3$ ITER Organization, Route de Vinon sur Verdon, 13067 St Paul Lez Durance Cedex, France}
\address{$^4$ University of Wisconsin, Madison, WI 53706-1609, USA}
\address{$^5$ Max Planck Institute for Plasma Physics, Boltzmannstr. 2, 85748 Garching, Germany}
\ead{i.krebs@differ.nl}
\begin{abstract}
A benchmark exercise for the modeling of vertical displacement events (VDEs) is presented and applied to the 3D nonlinear magneto-hydrodynamic codes M3D-C$^1$, JOREK and NIMROD. The simulations are based on a vertically unstable NSTX equilibrium enclosed by an axisymmetric resistive wall with rectangular cross section. A linear dependence of the linear VDE growth rates on the resistivity of the wall is recovered for sufficiently large wall conductivity and small temperatures in the open field line region. The benchmark results show good agreement between the VDE growth rates obtained from linear NIMROD and M3D-C$^1$ simulations as well as from the linear phase of axisymmetric nonlinear JOREK, NIMROD and M3D-C$^1$ simulations. Axisymmetric nonlinear simulations of a full VDE performed with the three codes are compared and excellent agreement is found regarding plasma location and plasma currents as well as eddy and halo currents in the wall.
%despite differences between the three codes, for example with respect to using a full or a reduced MHD model.

\vspace{0.5cm}
\noindent
The following article will be submitted to Physics of Plasmas. If accepted, it will be found at \url{https://aip.scitation.org/journal/php} after it is published.
\end{abstract}

\maketitle

\section{Introduction}

A vertical displacement event (VDE) denotes the vertical movement of a tokamak plasma toward the vessel walls, generally leading to a complete loss of the plasma confinement. In case of a \emph{cold VDE}, the loss of the control of the plasma position is caused by a rapid change of the plasma pressure and current density profile due to a thermal quench. In a \emph{hot VDE}, the plasma initially maintains most of its thermal energy and the loss of control occurs for different reasons, for example when a stability threshold in elongation is exceeded. The vessel wall currents produced by the VDE can lead to large transient forces on the vessel. Those forces have the potential to create significant mechanical stresses in the vacuum vessel, especially in large tokamaks. Therefore, comprehensive analysis, including large-scale simulations seeking to identify the worst case VDEs, are necessary as part of the design process.

Although a hot VDE is initially an axisymmetric instability, 3D instabilities develop during the course of the event. In particular, when the edge safety factor drops below some critical value due to scraping-off by the wall or by impurity cooling of the edge, the plasma becomes unstable to an external kink or resistive wall mode (RWM). These asymmetries can lead to asymmetric forces on the vessel walls. Since asymmetric forces will lead to asymmetric stresses, and might even rotate in mechanical resonance with the vessel structures \cite{Schioler2011}, they can lead to large local stresses in the vacuum vessel.

Employing 3D nonlinear magneto-hydrodynamic (MHD) codes to understand and predict the consequences of different types of VDEs for tokamak operation has become an active field of study. Recently, 3D VDE simulations have been performed by Strauss \cite{Strauss2015,Strauss2018} with the M3D code, Pfefferl{\'e} et al. \cite{Pfefferle2018} with M3D-C$^1$, Artola \cite{Artola2018} with JOREK and Sovinec et al. \cite{Sovinec2018} using NIMROD.

Combining a global 3D MHD model for the plasma evolution with implicit time stepping and a model for a resistive wall, NIMROD \cite{Sovinec2004}, JOREK \cite{Huysmans2007,Hoelzl2012} and M3D-C$^1$ \cite{Jardin2012,Ferraro2016} are among a small set of codes possessing the necessary capabilities for 3D VDE simulations. Nevertheless, a benchmark between any of the three codes involving VDE calculations had not been performed. In the following, we present the set up and results of such a benchmark exercise between M3D-C$^1$, NIMROD, and JOREK which is based on a vertically unstable NSTX equilibrium. Note that although an experimental equilibrium is used, this work is solely intended to be an inter-code benchmark exercise with the purpose of code verification focused on VDE relevant physics. An experimental validation of the physics models implemented is a separate, but equally important and ongoing endeavor. The goal of this work is to provide the community with a useful set of standard benchmark cases to be used to test codes which are to be applied to study VDEs.

The benchmark calculations presented in this paper are strictly 2D (axisymmetric) even though the three codes involved are fully 3D. We consider these 2D benchmarks to be an initial essential step in a series of benchmarks of increasing complexity leading to full 3D nonlinear simulations of different types of VDEs.

In Section~\ref{sec:vde}, a brief description of the M3D-C$^1$ model used for VDE calculations is given. In addition, some results on the influence of the temperature in the open field line region on VDE growth rates in M3D-C$^1$ simulations are discussed. The set up of the benchmark case is described in Section~\ref{sec:setup}. The differences between the three codes are discussed and the results of the benchmark are presented in Sections~\ref{sec:nim} - \ref{sec:jornl}. A summary and outlook on future work is given in Section~\ref{sec:sum}.

%\clearpage
\section{VDE simulations with M3D-C$^1$}
\label{sec:vde}
\begin{wrapfigure}{r}{0.5\textwidth}
\centering
	\includegraphics[width=0.49\textwidth]{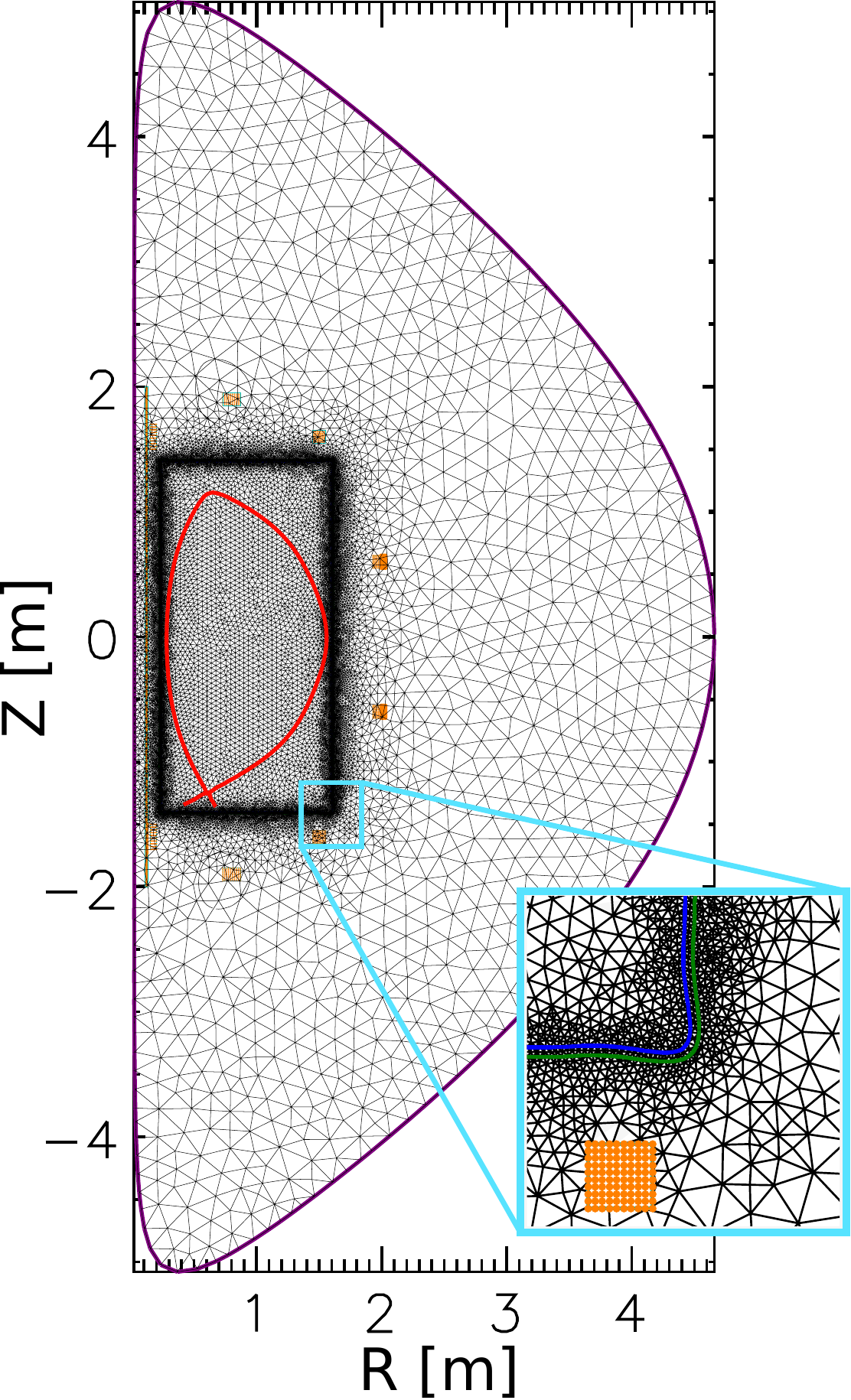}
	\caption{\label{fig:mesh}
	Mesh used for VDE benchmark case with M3D-C$^1$ (black). Shown are the ideal wall domain boundary (purple), the coils (orange), the thick resistive wall (between green and blue line) and the separatrix (red). The mesh has approximately 35000 elements.
	}
\end{wrapfigure}
The M3D-C$^1$ code is a high-order finite element code that solves the nonlinear time-dependent extended MHD equations. It uses a split-implicit time advance in order to enable simulations over transport time scales. For the spatial discretization, triangular  wedge  finite  elements are  used \cite{Jardin2010}.

For simulations of VDEs a three region model (as illustrated in Fig.~\ref{fig:mesh}) is used: Within the central region, the nonlinear extended MHD equations as described in \cite{Jardin2012} are solved. It is enclosed by a resistive wall region of arbitrary thickness. Between the resistive wall and the outer domain boundary, there is a vacuum region. The mesh resolution can be locally increased where needed, e.g. in the vicinity of the resistive wall.

At the boundary between the plasma domain and the resistive wall domain, no-slip boundary conditions are employed, and the temperature and density are kept at fixed values. Inside the resistive wall domain, the magnetic field is evolved according to $\partial_t \mathbf{B} = - \nabla \times (\eta_{\textnormal{wall}} \mathbf{j})$ where $\eta_{\textnormal{wall}}$ is the wall resistivity and the current density is given by $\mathbf{j} = (\nabla \times \mathbf{B})/\mu_0$. In the vacuum region $\mathbf{j} = 0$. There are no boundary conditions on the magnetic field at the resistive wall. Halo currents can flow into and out of the wall. Ideal wall boundary conditions are used at the outer domain boundary. The sensitivity of the VDE growth rates to the location of the ideal wall has been tested and the influence of the ideal wall has been estimated to result in a deviation below $10\%$ relative to no-outer-wall conditions.

The resistive wall model has been successfully benchmarked against analytic solutions for RWMs \cite{Ferraro2016}. The model has recently been extended to provide an option for a spatially varying resistivity which can be used to model conducting and non-conducting structures around the plasma. Furthermore, impurity and pellet models have been recently added to the plasma model for disruption mitigation studies \cite{Ferraro2018}. These options however have not been used for the calculations presented here.

M3D-C$^1$ can be used for 2D and 3D linear as well as for 2D axisymmetric nonlinear and 3D nonlinear simulations. 2D nonlinear simulations can also be restarted in both linear or 3D nonlinear mode.

\begin{figure}
	\centering
	\includegraphics[width=0.4\textwidth]{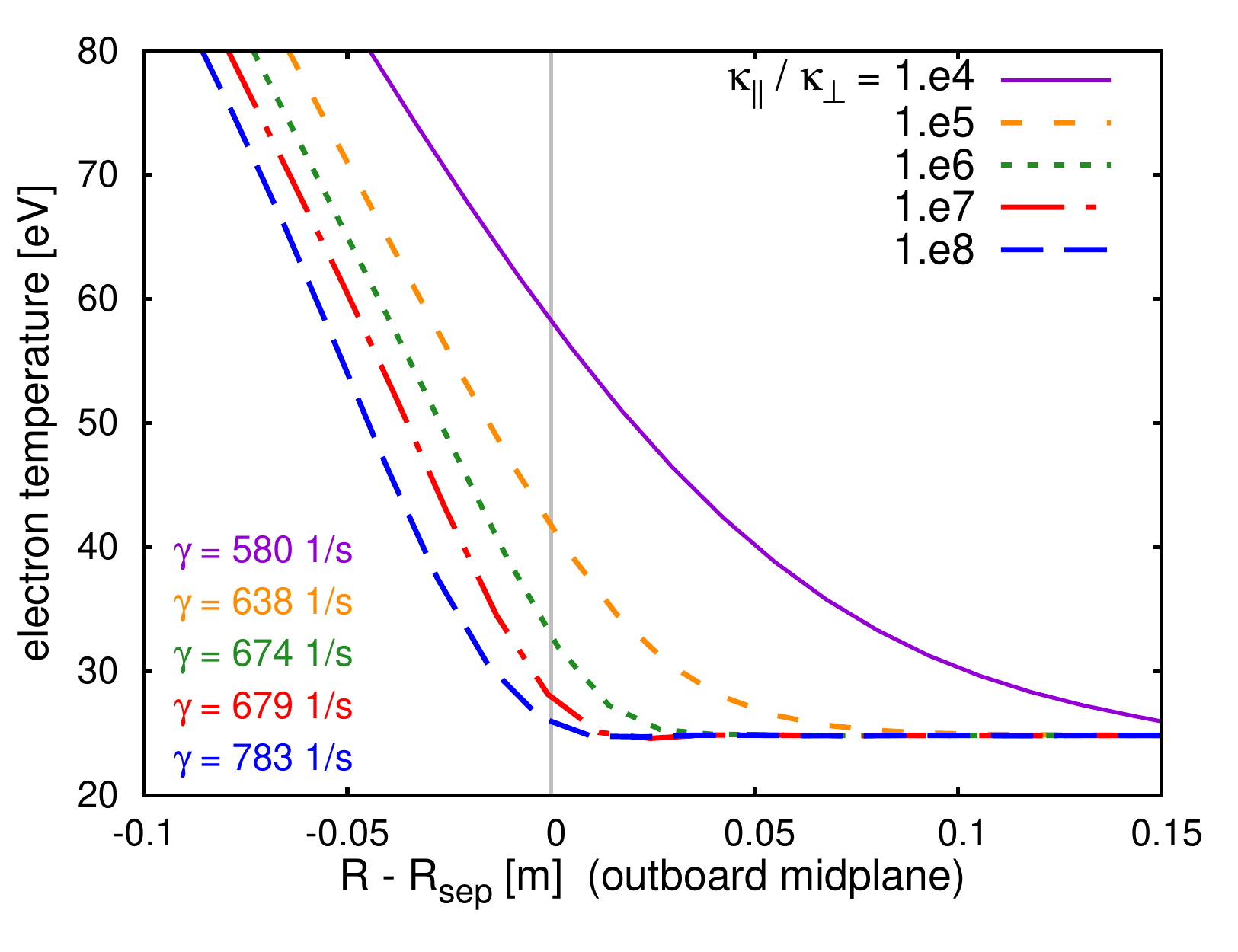}
	\caption{\label{fig:halowidth}
	Electron temperature at the outboard midplane versus the distance from the separatrix in a set of 2D nonlinear VDE simulations with different values of $\kappa_{\parallel}/\kappa_{\perp}$. VDE growth rates ($\gamma$) during the early drift phase have been obtained via an exponential fit to the time traces of $Z_{\textnormal{axis}}$ (during the initial $\SI{0.1}{\meter}$ of the displacement). Simulations are based on a DIII-D like equilibrium and have been performed with M3D-C$^1$.
	}
\end{figure}

In some axisymmetric codes that are used for VDE calculations such as DINA \cite{Khayrutdinov1993} and TSC \cite{Jardin1986}, the width of the halo region is an input parameter. Note that this is not the case in M3D-C$^1$, NIMROD and JOREK, where the width of the halo region is not artificially set, but is self-consistently determined by the heat transport model and can be adjusted via the heat diffusion anisotropy. Fig.~\ref{fig:halowidth} shows the resulting halo widths on the midplane for 2D nonlinear VDE simulations performed with M3D-C$^1$ with different values of the heat diffusion anisotropy, i.e. the ratio of the parallel heat diffusion coefficient $\kappa_\parallel$ to the perpendicular heat diffusion coefficient $\kappa_\perp$. (Here, the value of $\kappa_{\parallel}$ has been varied while keeping the value of $\kappa_{\perp}$ fixed.) Since response currents in the halo region slow down the VDE, smaller values of the heat diffusion anisotropy leading to higher edge temperatures cause reduced VDE growth rates ($\gamma$).

The temperatures in the open field line region, affected by both the edge temperature boundary condition and by the ratio of thermal conductivities as shown in Fig.~\ref{fig:halowidth}, can play an important role in VDE simulations. In the initial phase of hot VDEs, the vertical movement of the plasma is inhibited by response currents in the conducting vessel structures. Due to the finite resistivity of these structures, the response currents decay, and thus the linear VDE growth rate is determined by the resistive time of the vessel in this initial phase. However, if the electron temperatures in the open field line region are large in a simulation, its resistance can become sufficiently small to compete with the vessel resistance leading to response currents forming in the open field line region.

\begin{figure}
	\centering
	\includegraphics[width=0.9\textwidth]{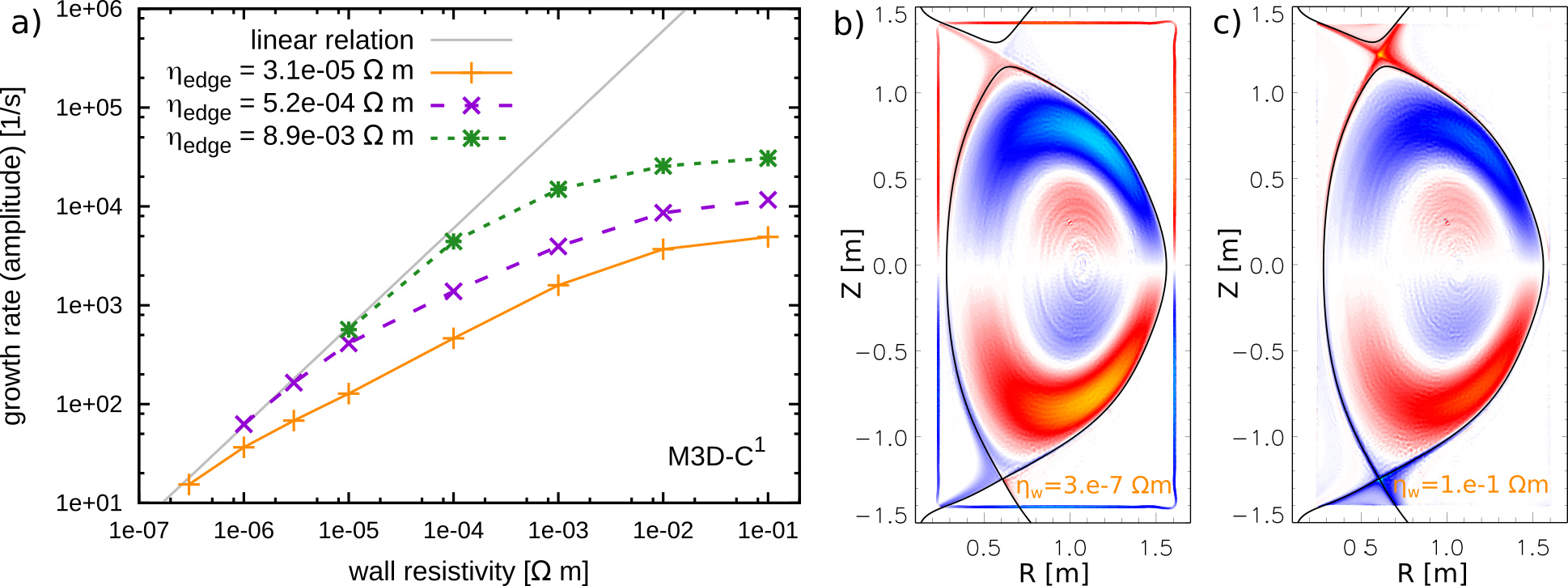}
	\caption{\label{fig:scan}
	a) Linear VDE growth rates obtained from M3D-C$^1$ simulations for different values of the wall resistivity ($\eta_{\textnormal{w}}$) and of the resistivity in the open field line region ($\eta_{\textnormal{edge}}$). Contour plots show the toroidal current density eigenfunctions of a case with $\eta_{\textnormal{edge}}=\SI{3.1e-5}{\ohm \meter}$, $\eta_{\textnormal{w}}=\SI{3.e-7}{\ohm \meter}$ where response currents form in the wall (b) and a case with $\eta_{\textnormal{edge}}=\SI{3.1e-5}{\ohm \meter}$, $\eta_{\textnormal{w}}=\SI{1.e-1}{\ohm \meter}$ where response currents form in the open field line region (c). (Red is positive, blue is negative, the separatrix is shown in black.)
	}
\end{figure}

Fig.~\ref{fig:scan}(a) shows the linear VDE growth rates for different values of the wall resistivity and different values of the edge temperature in linear M3D-C$^1$ simulations. By edge temperature we mean the value of the electron temperature at the boundary between the plasma domain and the resistive wall domain. This value is changed by changing the value of the equilibrium edge pressure while keeping the same edge density. Note that details on the set up and parameters of these simulations are given in Section~\ref{sec:setup} and Table~\ref{tab:param}(a).

In the limit of small wall resistivities and small edge temperatures for which the edge plasma resistivity is very large, the expected linear dependence of the growth rate on $\eta_{\textnormal{wall}}$ is recovered. For larger edge temperatures (i.e. lower $\eta_{\textnormal{edge}}$) or larger values of $\eta_{\textnormal{wall}}$, the resistance of the wall becomes comparable to the resistance of the open field line region. In this regime, the VDE growth is slowed down by response currents in the open field line region as illustrated in Figs.~\ref{fig:scan}(b) and (c). In the limit where the wall resistance is much larger than the resistance of the open field line region, the VDE growth rate is only determined by the plasma edge resistivity and becomes independent of the wall resistivity.

\section{Benchmark set up}
\label{sec:setup}
\begin{figure}
	\centering
	\includegraphics[width=0.2\textwidth]{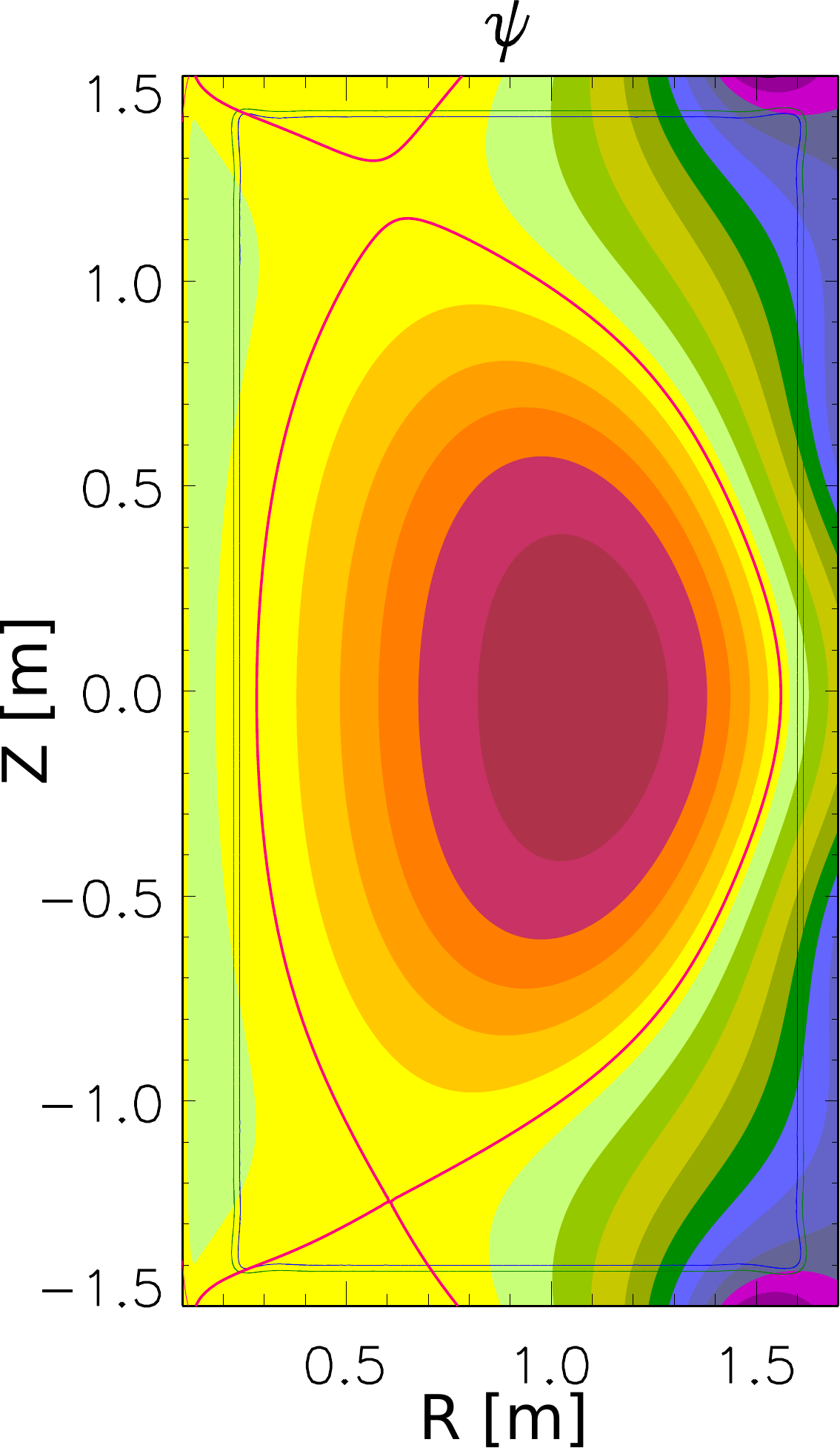}
	\caption{\label{fig:equi}
	Equilibrium poloidal magnetic flux of the VDE benchmark case (M3D-C$^1$). Also shown are the separatrix (red line) and the resistive wall (green and blue lines).
	}
\end{figure}
The equilibrium used for this benchmark case is loosely based on the NSTX discharge \#139536 at $t=\SI{309}{\milli \second}$. It is illustrated in Fig.~\ref{fig:equi}. An axisymmetric rectangular resistive wall is used to simplify the geometry. The corners of the inner boundary of the resistive wall domain are at ($R=\SI{0.24}{\meter}$, $Z=\SI{\pm 1.4}{\meter}$) and ($R=\SI{1.6}{\meter}$, $Z=\SI{\pm 1.4}{\meter}$). The thickness of the resistive wall is set to $\Delta_w=0.015\,\si{\meter}$.

The equilibrium position of the magnetic axis is ($R_{\textnormal{axis}}=\SI{1.07}{\meter}$, $Z_{\textnormal{axis}}=\SI{-0.015}{\meter}$). The toroidal magnetic field on axis is $B_{\textnormal{tor}} = 0.37\, \si{\tesla}$, the total toroidal plasma current is $I_{\textnormal{tot}} = 5.7\cdot 10^{5}\, \si{\ampere}$. The difference between the poloidal magnetic flux at the boundary and at the magnetic axis is $\Psi_{\textnormal{bnd}}-\Psi_{\textnormal{axis}}=\SI{-0.059}{\volt \second}$, where $\Psi=-\int B_{\textnormal{pol}}\, dA / 2 \pi$. The temperature profile is given by $T_e(\Psi) = 1 \si{\kilo \electronvolt} \cdot ( p(\Psi)/p_{\textnormal{axis}})^{0.6}$. The pressure and current density profiles are defined in the \emph{geqdsk} equilibrium file. Note that the geqdsk file, and files containing the coil positions and currents are available as supplementary material to this article.

Dynamic viscosity, perpendicular and parallel heat diffusion coefficients and the particle diffusion coefficient are constant in space and time. Their values are given in Table~\ref{tab:param}(a) for the simulations discussed in Section~\ref{sec:vde} and Table~\ref{tab:param}(b) for Sections~\ref{sec:nim}, \ref{sec:jorlin} and \ref{sec:jornl}. The plasma resistivity is given by the Spitzer model (i.e. $\eta(T_e) = 1.03\cdot 10^{-4} \cdot Z \cdot \ln{\Lambda} \cdot (T_{e}[\si{\electronvolt}])^{-3/2}\, \si{\ohm \meter}$, where $Z=1$, $\ln{\Lambda}=17$). The ion mass is set to twice the proton mass. A loop voltage is not applied. 

\begin{table}
\renewcommand{\arraystretch}{1.2}
\begin{center}
  \begin{tabular}{ | l |}
  \hline
  a) Section~\ref{sec:vde}                                                             \\ \hline \hline
  $\nu = 5.16\cdot 10^{-6}\, \si{\kilogram (\meter \second)^{-1}}$   \\
  $\kappa_{\perp}= 7.7\cdot 10^{19}\, \si{ (\meter \second)^{-1}}$ \\
  $\kappa_{\parallel} = \kappa_{\perp}$                              \\ 
  $D_n = 1.54 \si{\meter^2 \second^{-1}}$                            \\
  $T_{\textnormal{e,edge}} = 0.338\, \si{\electronvolt}$, $2.25\, \si{\electronvolt}$, $14.65\, \si{\electronvolt}$ \color{white}{..............................} \\
  \hline
  \end{tabular}
  
  \vspace{0.4cm}
    \begin{tabular}{ | l |}
  \hline
  b) Sections~\ref{sec:nim},\ref{sec:jorlin},\ref{sec:jornl} \\ \hline \hline
  $\nu = 5.16\cdot 10^{-7}\, \si{\kilogram (\meter \second)^{-1}}$ \\
  $\kappa_{\perp}= 1.54\cdot 10^{18}\, \si{ (\meter \second)^{-1}}$  \\
  Section~\ref{sec:nim}: $\kappa_{\parallel} = \kappa_{\perp}$; Section~\ref{sec:jorlin}\&\ref{sec:jornl}: $\kappa_{\parallel} = \kappa_{\perp} \cdot 10^5$ \\ 
  $D_n = 1.54\cdot 10^{-1} \si{\meter^2 \second^{-1}}$ \\
  Section~\ref{sec:nim}\&\ref{sec:jornl}: $T_{\textnormal{e,edge}} = 14.65\, \si{\electronvolt}$; Section~\ref{sec:jorlin}: $T_{\textnormal{e,eff}} = 1.\, \si{\electronvolt}$ \\ 
  \hline
  \end{tabular}
\caption{Dynamic viscosity $\nu$, perpendicular and parallel heat diffusion coefficients, $\kappa_{\perp}$ and $\kappa_{\parallel}$, particle diffusion coefficients $D_n$ and electron temperatures at the boundary to the wall $T_{\textnormal{e,edge}}$ used for the different sets of simulations. $T_{\textnormal{e,eff}}$ is defined as $T_{\textnormal{e,edge}}-T_{e,\textnormal{off}}$ where $T_{e,\textnormal{off}}$ is an offset temperature (see Section~\ref{sec:jorlin}).}
\label{tab:param}
\end{center}
\renewcommand{\arraystretch}{1.}
\end{table}

\section{NIMROD \& M3D-C$^1$ - linear simulations}
\label{sec:nim}
\begin{figure}
	\centering
	\includegraphics[width=0.5\textwidth]{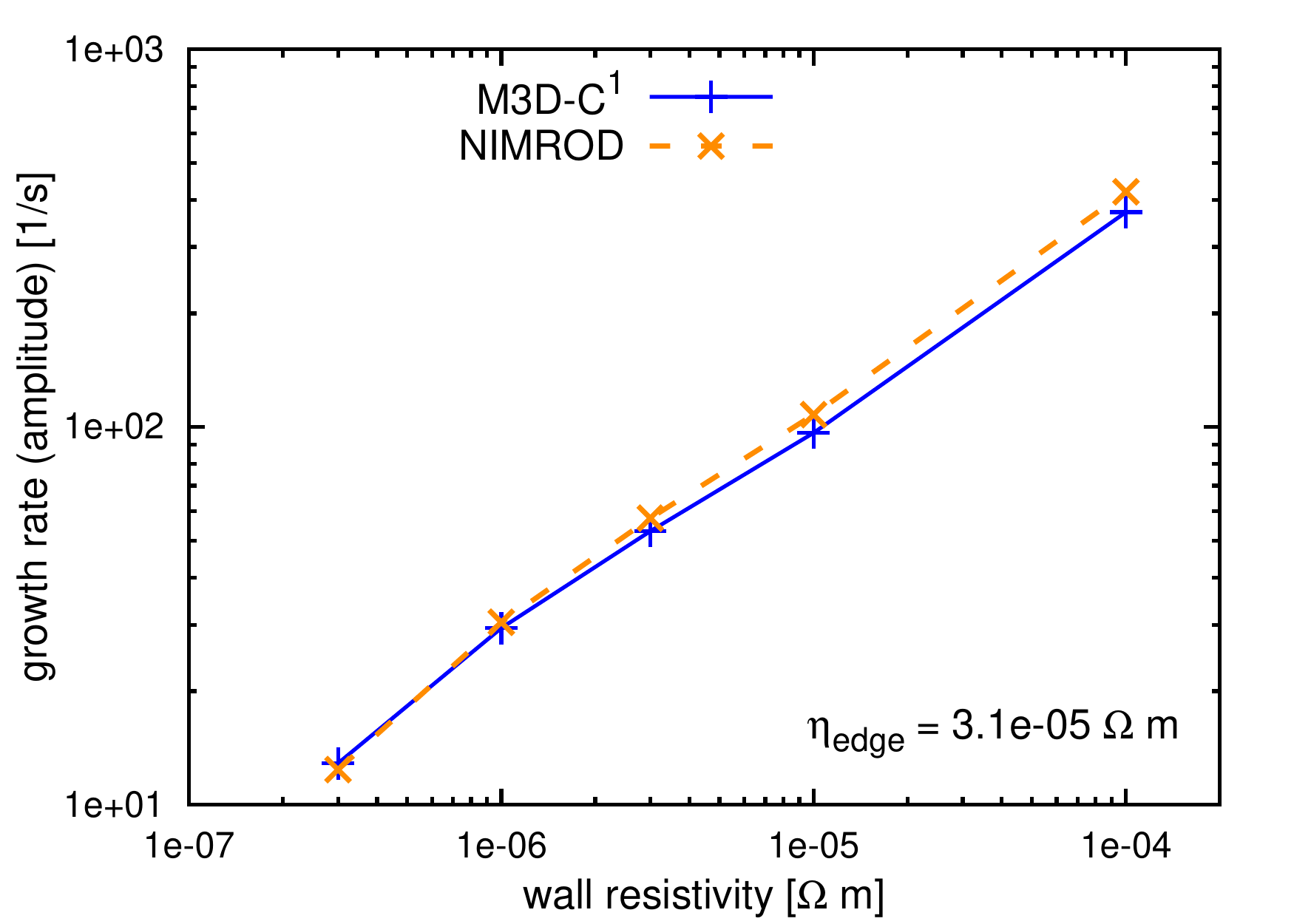}
	\caption{\label{fig:nim}
	Comparison of VDE growth rates from linear M3D-C$^1$ and NIMROD simulations. The growth rates deviate by between $4\%$ and $13\%$.
	}
\end{figure}

While NIMROD and M3D-C$^1$ have similar physics models, the numerical methods differ, which makes benchmarks between these two codes particularly valuable. In contrast to M3D-C$^1$, NIMROD uses high-order $C^0$ quadrilateral finite elements in the $R$-$Z$ plane and a Fourier spectral representation for the toroidal discretization. When testing resolution for these computations, bicubic and biquartic elements have been applied. As described in Ref.~\cite{Sovinec2018}, the NIMROD computations presented here use a thin-wall model that couples dynamics in the plasma subdomain with numerically computed magnetic responses in an outer vacuum subdomain.  This differs from M3D-C$^1$, which represents a thick wall within a single mesh that spans all domains.

The geometry of the outer subdomain used in the NIMROD benchmark computations matches the shape shown in Fig.~\ref{fig:mesh}, except that the top and bottom corners at $R=\SI{0.02}{\meter}$ are not rounded.  The computations also used fixed viscosity and thermal conductivity coefficients (same as M3D-C$^1$) without the dependence on plasma density that is shown in Refs.~\cite{Sovinec2018,Sovinec2004}.

Fig.~\ref{fig:nim} shows a comparison of the linear VDE growth rates obtained from linear M3D-C$^1$ and NIMROD simulations (using the parameters listed in Table~\ref{tab:param}(b)). The growth rates agree well over a wide range of wall resistivities. The largest deviation between the growth rates is $13\%$ and it occurs at the largest values of the wall resistivity, where the results are most sensitive to the representation of the equilibrium and halo responses.

\section{JOREK, NIMROD \& M3D-C$^1$ - linear phase of axisymmetric nonlinear simulations}
\label{sec:jorlin}
\begin{figure}
	\centering
	\includegraphics[width=0.5\textwidth]{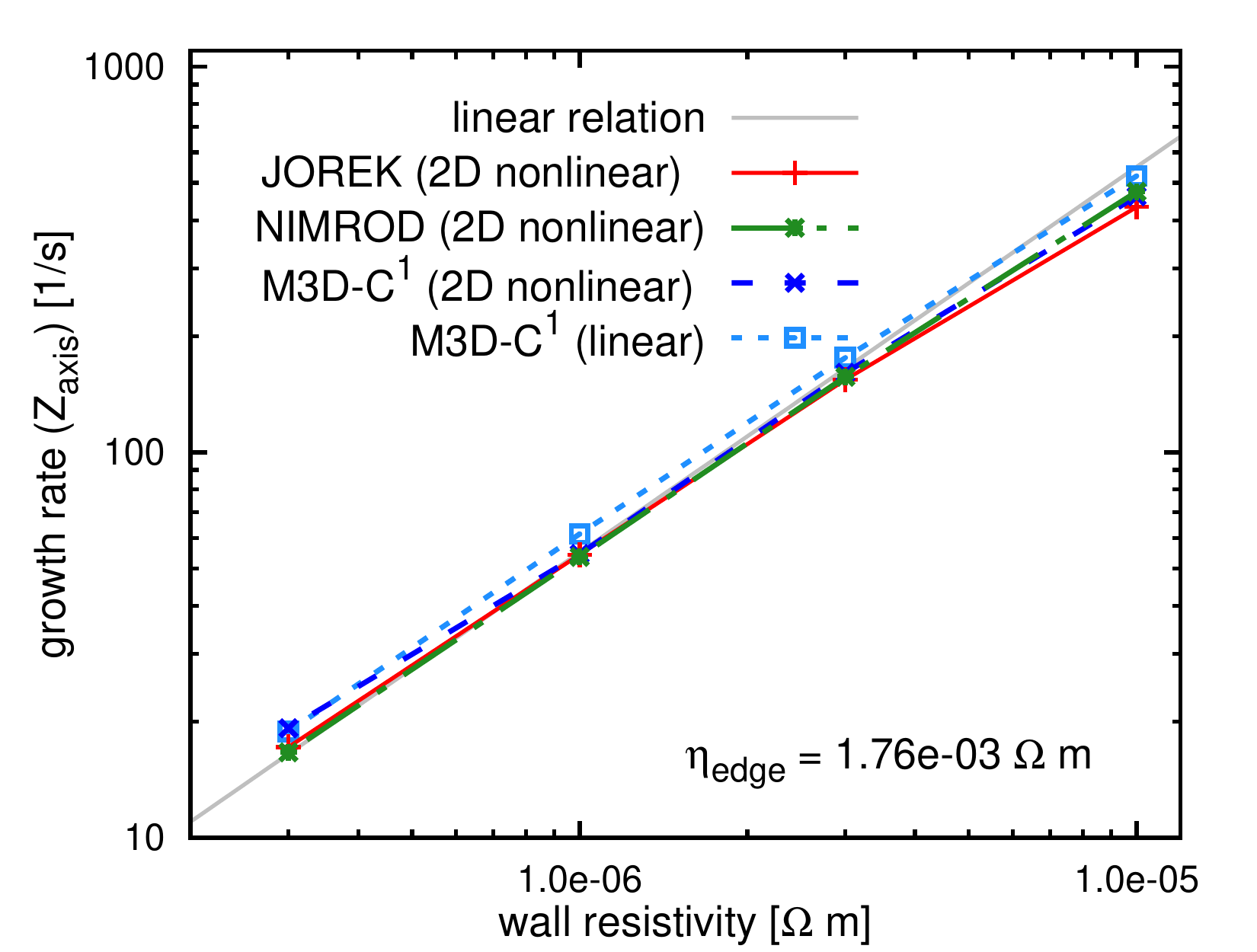}
	\caption{\label{fig:jorlin}
	Comparison of VDE growth rates from the linear phase of 2D nonlinear M3D-C$^1$, NIMROD and JOREK simulations. They deviate between $0.3\%$ and $15\%$. Also shown are the results of linear M3D-C$^1$ calculations.
}
\end{figure}

For simulations with JOREK that include a resistive wall, the JOREK-STARWALL coupling is used \cite{Hoelzl2012,Merkel2015}. Similar to NIMROD and in contrast to M3D-C$^1$, JOREK uses a spectral representation for the toroidal discretization. Cubic B{\'e}zier finite elements are used for the discretization in the $R$-$Z$ plane. There are a few differences between the model that JOREK uses for the benchmark simulations and the models that M3D-C$^1$ and NIMROD use: (i) Although JOREK has a full MHD model, it uses a reduced MHD model \cite{Huysmans2009} for the VDE calculations presented here since the JOREK-STARWALL coupling is not yet available for the full MHD model. (ii) In JOREK-STARWALL, the vacuum  contribution is implemented by using a Green's function method. Therefore, it is not necessary to discretize the vacuum region and apply ideal wall boundary conditions in an outer boundary. This property comes from the fact that the full vacuum response can be expressed as a function of the magnetic field at the plasma boundary. (iii) At the resistive wall, instead of no-slip boundary conditions, only the normal component of the velocity vanishes.

For the JOREK simulations presented here, a polar grid is used with increased resolution in the region surrounding the point of contact between plasma and wall. The number of B{\'e}zier elements used is 22000 and the number of linear triangular elements used for the representation of the wall is 48000.

Since JOREK does not have an option that allows linear simulations with toroidal mode number $n=0$, we compare the VDE growth rates in the early, linear phase of the evolution obtained in 2D axisymmetric nonlinear simulations.

In order to be able to run benchmark cases in the regime where the VDE growth rate is not influenced by response currents in the open field line region, the value of the edge temperature has to be sufficiently small. Since in nonlinear simulations too small values of the edge temperature can lead to numerical problems, we use a small temperature offset only within the calculation of the Spitzer resistivity such that $\eta(T_e) = \eta_{\textnormal{Spitzer}}(T_e - T_{e,\textnormal{off}})$ in all three codes. Here, the edge temperature is $T_{e,\textnormal{edge}} = \SI{14.65}{\electronvolt}$ and the offset is $T_{e,\textnormal{off}} = \SI{13.65}{\electronvolt}$ which results in an effective edge resistivity corresponding to a temperature of $T_{e,\textnormal{eff}} = \SI{1}{\electronvolt}$. For simplicity, the Ohmic heating term in the temperature equation is switched off. 

Figure~\ref{fig:jorlin} shows a comparison of the resulting VDE growth rates. They have been obtained from the 2D nonlinear simulations by fitting $Z_{\textnormal{axis}} = a + b \cdot \exp(\gamma t)$ to the time trace of the vertical position of the magnetic axis where $\gamma$ is the growth rate. Only the early, linear phase of the evolution (vertical position of magnetic axis between $Z_{\textnormal{axis}} = \SI{-1.64}{\centi \meter}$ and $Z_{\textnormal{axis}} = \SI{-3.04}{\centi \meter}$) has been taken into account.

All three codes find the expected linear relation between VDE growth rate and wall resistivity and the results agree well. The deviation between the obtained growth rates is around $3\%$ or less for most wall resistivities and does not exceed $12\%$ in the other cases, except for a deviation of $15\%$ between the M3D-C$^1$ and the NIMROD result for the smallest wall resistivity. Also the growth rates obtained from linear M3D-C$^1$ simulations agree well with the results from the early phase of the 2D nonlinear simulations. However, it should be noted that such good agreement between the growth rates obtained from the linear and the nonlinear simulations is only achieved when the simulation is first run in 2D nonlinear mode for a few time steps (in this case until the plasma has drifted by $\sim \SI{2}{\milli \meter}$), and then restarted as a linear simulation. A possible reason for this might be small inconsistencies in the geqdsk equilibrium that relax quickly when run in nonlinear mode. The exact cause for this will need some further investigation.

\section{JOREK, NIMROD \& M3D-C$^1$ - axisymmetric nonlinear simulation}
\label{sec:jornl}
\begin{figure}
	\centering
	\includegraphics[width=0.8\textwidth]{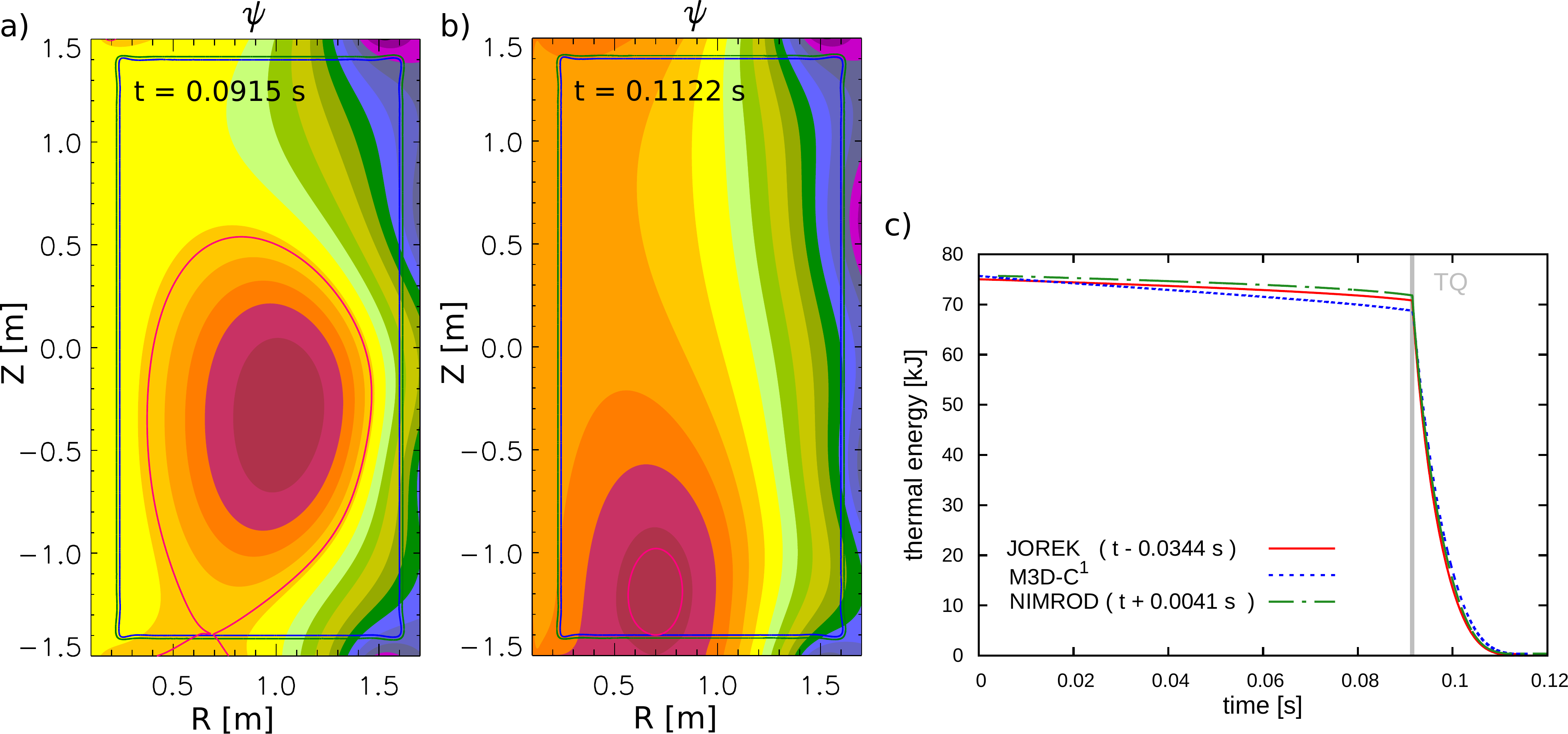}
	\caption{\label{fig:tq}
	Contour plots show the poloidal magnetic flux in the 2D nonlinear M3D-C$^1$ simulation at the point in time when the plasma first becomes limited by the wall (a) and close to the end of the VDE (b). The time traces of the thermal energy (c) in the M3D-C$^1$, NIMROD and JOREK simulation show the artificial thermal quench initiated when the plasma touches the wall ($t=\SI{0.0915}{\second}$ for M3D-C$^1$; JOREK and NIMROD traces have been shifted in time such that the points in time of the first plasma-wall contact coincide).
	}
\end{figure}

\begin{figure}
	\centering
	\includegraphics[width=0.9\textwidth]{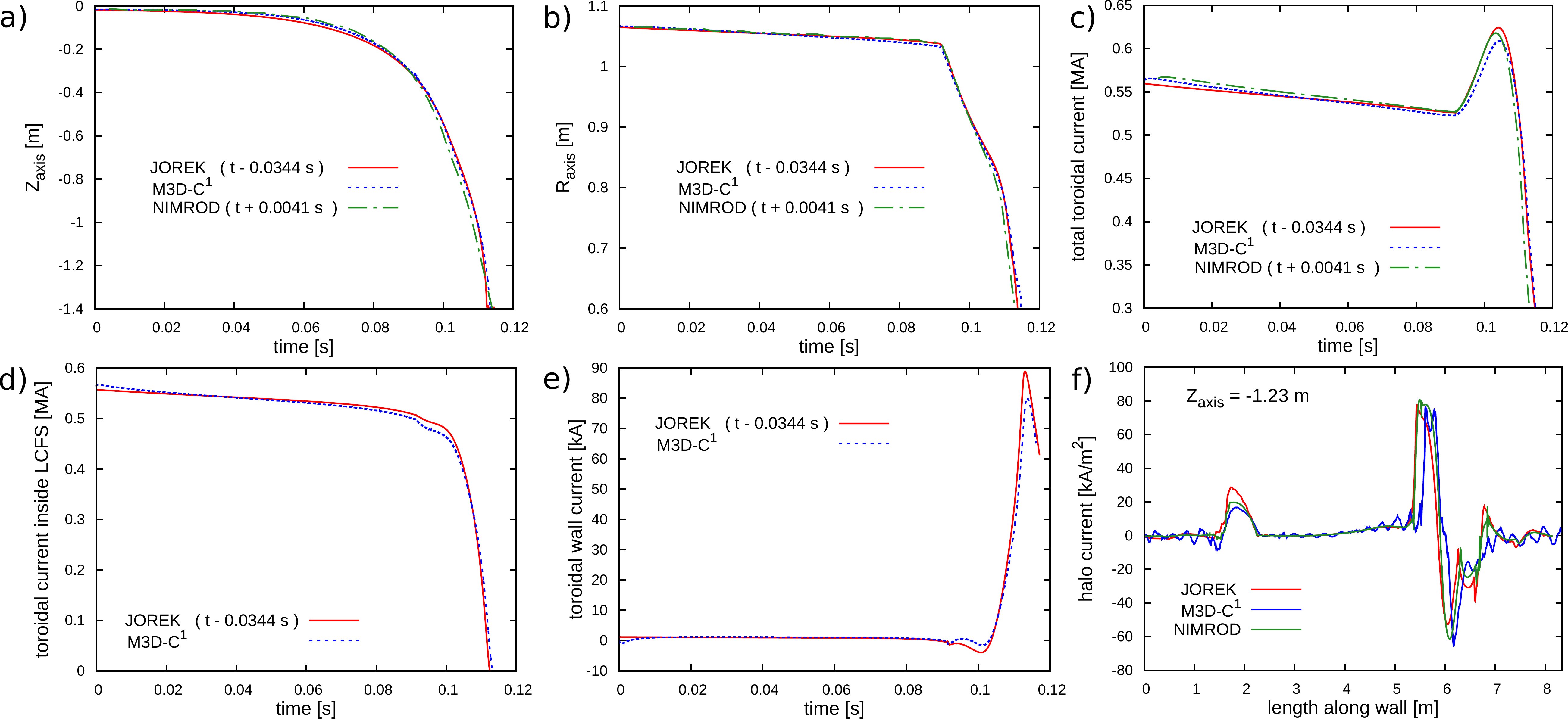}
	\caption{\label{fig:jornl}
	Comparison of time traces from a 2D nonlinear simulation performed with JOREK, NIMROD and M3D-C$^1$: a) vertical position of magnetic axis, b) radial position of magnetic axis, c) toroidal current inside the LCFS and the open field line region, d) toroidal current inside the LCFS, e) net toroidal wall current. JOREK and NIMROD time traces are shifted so that the points in time of the first plasma-wall contact coincide. f) shows the component of the current density that is normal to the wall traced along the length along the wall at the point in time when $Z_{\textnormal{axis}}=\SI{-1.23}{\meter}$. The trace starts at the low-field side midplane and continues counterclockwise.
	}
\end{figure}

In the following, the results obtained by JOREK, NIMROD and M3D-C$^1$ on the further axisymmetric nonlinear evolution of a VDE are compared. The set up and parameters of these simulations are the same as for the simulations discussed in Section~\ref{sec:jorlin}, except that $T_{e,\textnormal{off}}=0$ such that the edge resistivity corresponds to an edge electron temperature of $T_{e,\textnormal{edge}} = \SI{14.65}{\electronvolt}$. The resistivity of the wall has been set to $\eta_{\textnormal{w}} = 3. \cdot 10^{-6}\, \si{\ohm \meter}$. 

In addition, a thermal quench has been artificially initiated during the course of the evolution. In 3D nonlinear MHD simulations, e.g.~\cite{Pfefferle2018}, the decrease of the edge safety factor during the course of a VDE causes non-axisymmetric instabilities to develop. These 3D instabilities cause the magnetic flux surfaces to break up which leads to greatly increased thermal transport. Since this effect cannot occur in axisymmetric simulations, an artificial thermal quench is initiated by increasing the perpendicular heat diffusion coefficient by a factor of 500 when the plasma becomes limited by the wall. Also, the particle diffusion coefficient is multiplied by a factor of 20.

The poloidal magnetic flux at the point in time when the plasma becomes limited by the wall in the M3D-C$^1$ simulation and the time traces of the thermal energy in the M3D-C$^1$, JOREK and NIMROD simulation are shown in Fig.~\ref{fig:tq}.

In order to enable a meaningful comparison of the results, the signals are slightly shifted in time such that the points in time of the first plasma-wall contact, e.g. the start of the thermal quench, coincide. This compensates for differences caused by the exponential dependency on the initial conditions. (The plasma first touches the wall at $t\approx \SI{126}{\milli \second}$ in the JOREK simulation, at $t\approx \SI{87.4}{\milli \second}$ in the NIMROD simulation and at $t\approx \SI{91.5}{\milli \second}$ in the M3D-C$^1$ simulation.) Fig.~\ref{fig:jornl} compares the time traces of the vertical and radial positions of the magnetic axis, the toroidal current enclosed by the last closed flux surface (LCFS), the total toroidal current inside the LCFS and in the open field line region and the net toroidal current in the resistive wall. 

In addition, the halo current at the plasma-wall interface, i.e. the component of the current density perpendicular to the wall at the wall, is shown for a point in time during the late evolution (when $Z_{\textnormal{axis}}\approx \SI{-1.23}{\meter}$). The halo current is plotted against the distance along the wall, measured counter-clockwise, starting at the low-field side midplane. For the JOREK simulation, the halo current is calculated from $\mathbf{j} \times \mathbf{B} = \nabla p$, assuming that the plasma is in equilibrium. Note that the location of the halo current spikes resulting from the M3D-C$^1$ simulation appears to be slightly shifted with respect to the other two traces. This is an artefact caused by the M3D-C$^1$ resistive wall having a slightly larger circumference since its corners are less rounded then the ones of the resistive walls used for the JOREK and NIMROD simulations.

As expected, the halo current flows into and out of the wall in a narrow region surrounding the contact point of the last closed flux surface and the wall. Despite the differences in physics models and numerical implementation between the three codes, the results agree well.

\section{Summary \& Outlook}
\label{sec:sum}

A VDE benchmark case for nonlinear MHD codes has been presented. It is based on a vertically unstable NSTX equilibrium and uses an axisymmetric rectangular model for the resistive wall. The 3D nonlinear MHD codes M3D-C$^1$, NIMROD and JOREK are applied to the benchmark case, but run in a linear as well as a 2D (axisymmetric) nonlinear mode. Linear simulations show that the expected linear dependence of the VDE growth rate on the resistivity of the vessel wall ($\eta_{\textnormal{w}}$) is recovered for small values of $\eta_{\textnormal{w}}$ and a sufficiently large resistivity in the open field line region. If the temperature in the open field line region becomes too large, response currents build in this region and slow down the VDE growth.

Agreement within approximately $10\%$ is found between the results of the linear benchmark simulations obtained by NIMROD and M3D-C$^1$, as well as between the VDE growth rates in the linear phase of axisymmetric nonlinear simulations performed with NIMROD, M3D-C$^1$ and JOREK.

The further axisymmetric nonlinear evolution of a selected case has been calculated using the three codes and the time traces of the position of the magnetic axis, the toroidal plasma current and the toroidal current in the resistive wall as well as the resulting halo current have been compared. Despite the differences in physics models and numerical methods, the results agree well. 

In particular, this implies that for the benchmark cases presented here the reduced MHD model that JOREK employs reproduces the results of the full MHD models of NIMROD and M3D-C$^1$. The good agreement between the reduced MHD model and the full MHD models originates from the presence of a large vacuum toroidal field \cite{Wigger2011}. The formulation of the energy principle for $n=0$ reveals that important stabilizing terms involving the large toroidal magnetic field ($B_\phi$) can be minimized by choosing the following form of the plasma velocity $\mathbf{v}$ (or plasma displacement $\mathbf{\xi}$):
\begin{align}
\mathbf{v}=\mathbf{\xi} \gamma = -R^2 \nabla u \times \nabla \phi
\end{align}

where  $\gamma$ is the growth rate, $\phi$ is the toroidal angle and $u$ is the velocity stream function. This condition is called ``Slip Motion Condition'' \cite{Rebhan1976}, in which the plasma moves across the large toroidal field without doing work against it.

The velocity representation of the ``Slip Motion Condition'' is the main assumption used in reduced MHD, which is equivalently justified whenever $F\equiv RB_\phi \approx F_\textnormal{vacuum}$ and explains the excellent agreement with the full MHD models. The nature of the problem makes the full velocity to be well represented by the reduced MHD velocity. In the presented simulations the total toroidal field  differed from the vacuum toroidal field by a maximum of $5\%$ despite the small aspect ratio of the configuration.

Future work will include an extension of this benchmark towards fully 3D nonlinear simulations with the three codes. Furthermore, we hope that other linear, axisymmetric nonlinear, or 3D nonlinear codes used for VDE calculations might be applied to the benchmark case presented here as well.

\section*{Supplementary Material}

See supplementary material for the geqdsk file defining the equilibrium, and files containing the coil positions and currents for the discussed benchmark cases.

\ack
This work was supported by US DOE Contracts No.~DE-AC02-09CH11466 and No.~DE-SC0018001, and the SciDAC Center for Tokamak Transient Simulations. Some of the computations presented in this article used resources of the National Energy Research Scientific Computing Center (NERSC), a U.S. Department of Energy Office of Science User Facility operated under Contract No.~DE-AC02-05CH11231. The  support  from  the  EUROfusion  Researcher  Fellowship  programme  under  the  task  agreement  WP19-20-ERG-DIFFER/Krebs is gratefully acknowledged.

\section*{Disclaimer}

ITER is the Nuclear Facility INB no.~174. The views and opinions expressed herein do not necessarily reflect those of the ITER Organization. This publication is provided for scientific purposes only. Its content should not be considered as commitments from the ITER Organization as a nuclear operator in the frame of the licensing process.

\linespread{0.9}
\section*{References}
\bibliographystyle{ieeetr}%iaeaconf
\bibliography{database.bib}
\end{document}